\documentclass[journal]{IEEEtran}

\usepackage{setspace}
\usepackage{cite,graphicx,amsmath,amssymb}
\usepackage{hyperref}
\usepackage{subfigure}
\usepackage{url}
\usepackage{fancyhdr}
\usepackage{mdwmath}
\usepackage{mdwtab}
\usepackage{balance}
\usepackage{xcolor}
\usepackage{bm}
\usepackage{amsthm}
\usepackage{threeparttable}
\usepackage{algorithm}
\usepackage{algorithmic}
\usepackage{multirow}
\usepackage{flafter}
\usepackage{makecell}
\usepackage{diagbox}
\usepackage{tcolorbox}
\theoremstyle{definition}

\newtheorem{theorem}{Theorem}

\newtheorem{lemma}{Lemma}

\newtheorem{corollary}{Corollary}

\usepackage{xurl}

\providecommand{\url}[1]{#1}
\allowdisplaybreaks
\begin{document}

\title{Simultaneously Transmitting And Reflecting Surfaces (STARS) for Multi-Functional 6G}

\author{Xidong Mu, Zhaolin Wang, and Yuanwei Liu,~\IEEEmembership{Fellow,~IEEE}\\

\thanks{Xidong Mu is with the Centre for Wireless Innovation (CWI), Queen's University Belfast, Belfast, BT3 9DT, U.K. (e-mail: x.mu@qub.ac.uk).}
\thanks{Zhaolin Wang is with the School of Electronic Engineering and Computer Science, Queen Mary University of London, London E1 4NS, U.K. (e-mail: zhaolin.wang@qmul.ac.uk).}
\thanks{Yuanwei Liu is with the School of Electronic Engineering and Computer Science, Queen Mary University of London (QMUL), E1 4NS London, U.K., and also with the Department of Electronic Engineering, Kyung Hee University, Yongin-si, Gyeonggi-do 17104, South Korea (e-mail: yuanwei.liu@ qmul.ac.uk)}
}

\maketitle
\begin{abstract}
Simultaneously transmitting and reflecting surface (STARS) empowered multi-functional 6G wireless networks are investigated. Starting with the communication functionality, various types of STARS are introduced in terms of power amplification capabilities, reciprocity features, and spatial density of elements. Then, three STARS-empowered wireless sensing architectures are proposed, namely STARS-aided monostatic sensing, STARS-enabled bistatic sensing, and sensing with target-mounted STARS, where the representative benefits and application challenges are identified. Furthermore, promising applications of STARS for computing and caching functionalities are explored to improve the computation efficiency and reduce the content delivery latency. Finally, recent standardization progress for reconfigurable intelligent surfaces is presented for motivating the employment of STARS in multi-functional 6G.  
\end{abstract}

\section{Introduction}
With the rapid worldwide commercialization of fifth-generation (5G) wireless networks, the development of sixth-generation (6G) wireless networks has become a focal point of academia and industries. It can be foreseen that compared to 5G, 6G will not only need to support ultra-high data rates and ultra-reliable connections but also enable multiple functionalities, such as sensing, localization, and computing, with a high degree of integration, namely multi-functional 6G~\cite{10529727}. This is essential for enabling new and revolutionary applications like autonomous driving, extended reality (XR), and telehealth. To achieve these ambitious goals, the developing trend of wireless technologies is to employ large-scale antenna arrays (e.g., extremely large multiple-input multiple-output (MIMO)) and ultra-high operating frequencies (e.g., terahertz (THz) communications). This, however, causes unaffordable high hardware costs and energy consumption due to the massive expensive components employed at transceivers.

Thanks to recent breakthroughs in metamaterials and fabrication technologies, reconfigurable intelligent surfaces (RISs) have become a promising cost- and energy-efficient technology for building 6G~\cite{9140329}. In contrast to conventional active and costly multiple-antenna arrays, RISs generally have a thin surface, on which a large number of low-cost and low-energy consumption electromagnetic (EM) elements are embedded. By adjusting the EM features of these elements, RISs can adjust the propagation of incident wireless signals and thus redirect them for signal enhancement and/or interference mitigation~\cite{9140329}. The key benefit is that RISs do not produce new wireless signals and only beneficially \emph{reuse} existing wireless signals to enhance the wireless coverage and quality with the facilitated `smart radio environment'. As a result, RISs would play a vital role in achieving sustainable 6G wireless networks. 

With the extensive research efforts devoted to RISs, diverse variants with different functionalities have been proposed in recent years. Among them, a novel concept of simultaneously transmitting and reflecting surfaces (STARS) was proposed in \cite{9570143}. Compared to existing RIS technologies which only reflect or transmit (refract) wireless signals, STARS integrates the two fundamental functionalities into the single metasurface. By doing so, the incident wireless signals can be transmitted and reflected to both sides. The unique advantages of STARS in the family of RIS technologies can be summarized as follows.
\begin{itemize}	
	\item \textbf{360° Smart Radio Environment}: The most prominent feature of STARS is that the 180° smart radio environment realized by conventional reflective/transmissive RISs is extended to a 360° one.  This groundbreaking enhancement enables all surrounding wireless devices to benefit from STARS, irrespective of their positions. 
    \item \textbf{Enhanced Design Degrees-of-Freedom (DoFs)}: STARS introduces adjustable transmission and reflection beamforming to reconfigure the propagation of wireless signals on both sides. Therefore, STARS provides enhanced DoFs for wireless designs over conventional reflective/transmissive RISs.
    \item \textbf{High Deployment Flexibility}: The full-space coverage of STARS enables versatile deployment strategies in wireless networks. Unlike reflective/transmissive RISs, which require precise orientations during deployment, STARS dynamically adapts to transmit or reflect incident wireless signals. This flexibility opens up a wide range of deployment locations for STARS.
\end{itemize}	
\begin{figure*}[t!]
\begin{center}
    \includegraphics[width=5in]{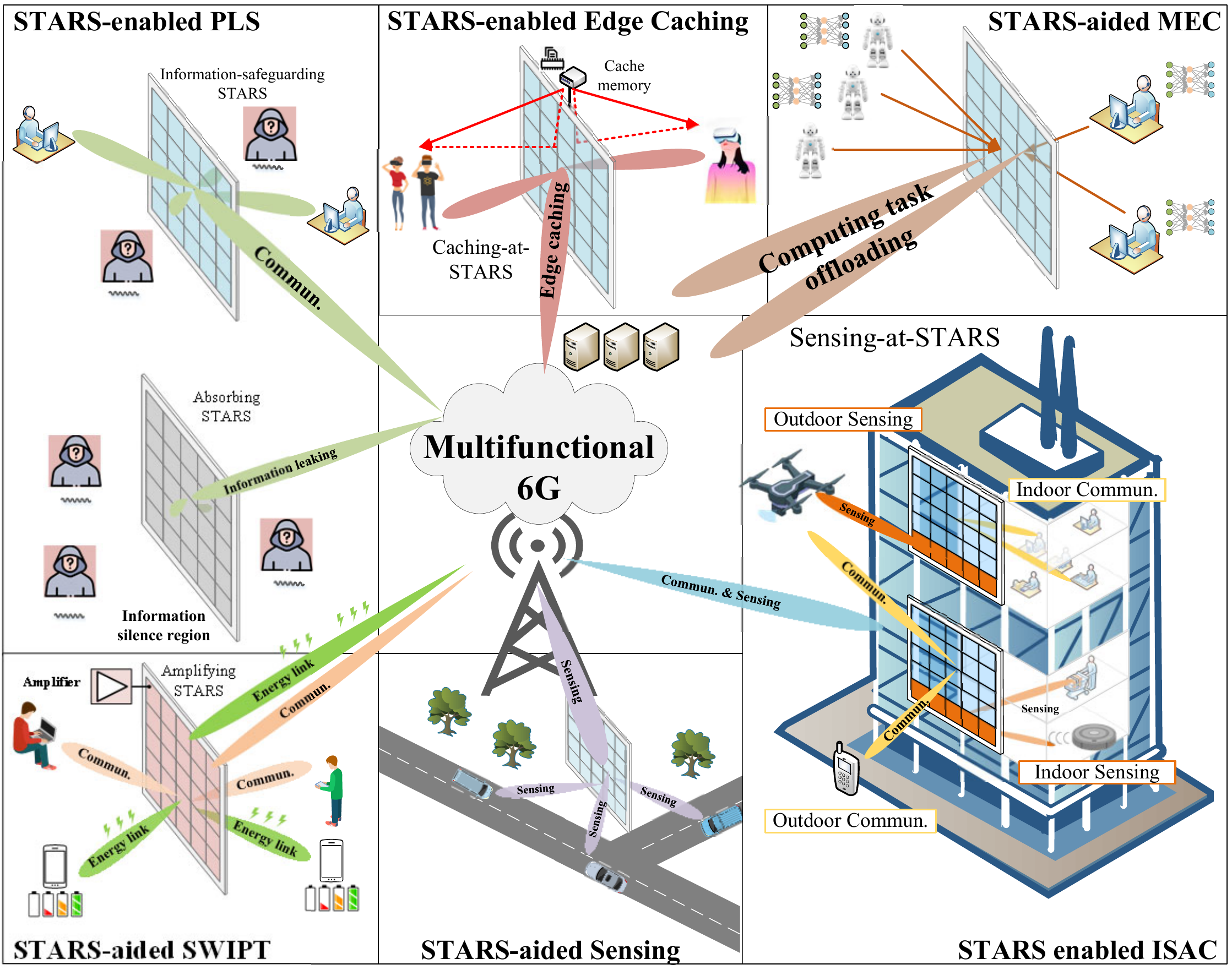}
    \caption{Applications of STARS in multi-functional 6G.}
    \label{STAR_1}
\end{center}
\end{figure*}

With the compelling benefits outlined above, STARS emerges as one of the most promising candidates for satisfying stringent communication requirements and seamlessly integrating multiple functionalities in 6G. Fig. \ref{STAR_1} illustrates the applications of different types of STARS in multi-functional 6G wireless networks. On the one hand, passive STARS can be seamlessly deployed to enhance wireless performance, such as physical layer security (PLS), mobile edge computing (MEC), and sensing coverage. On the other hand, active STARS with advanced functionalities can be further employed in different application scenarios, such as simultaneous wireless information and power transfer (SWIPT) and integrated sensing and communication (ISAC). For example, active STARS with the amplifying function can help to overcome severe path loss and thus improve the energy charging efficiency for SWIPT, while absorbing STARS has the potential to manually create a signal dead zone for anti-eavesdropping. As a further advance, active STARS with sensing/caching capabilities can be used to build ubiquitous perceptive wireless networks and reduce content delivery latency. 

Despite STARS exhibiting great potential in multi-functional 6G, existing research contributions mainly focused on the employment of STARS in wireless communications~\cite{9570143}. The investigation of STARS in other wireless functionalities is still in an early stage and its effectiveness has not been widely recognized. This provides our main motivation for developing this article to provide a systematic overview of STARS and promising use cases in multi-functional 6G. The main contributions of this article can be summarized as follows:
\begin{itemize}	
	\item We classify STARS employed in wireless communications based on their power amplification capability, reciprocity, and spatial density of elements. For each category, we discuss their representative advantages and disadvantages. 
    \item We investigate STARS for the sensing functionality, where several sensing architectures, including STARS-aided monostatic, STARS-enabled bistatic sensing, and sensing with target-mounted STARS, are presented along with their unique benefits and application challenges.
    \item We explore STARS for both computing and caching functionalities. In context of computing, promising applications of STARS in MEC and over-the-air computation (AirComp) are discussed to enhance the computation efficiency. Regarding caching, a caching-at-STARS architecture is put forward to mitigate content delivery latency.
    \item We review the state-of-the-art standardization activities for RISs and discuss the potential impact on advancing STARS-aided multi-functional 6G.
\end{itemize}

\begin{figure*}[t!]
\begin{center}
    \includegraphics[width=5in]{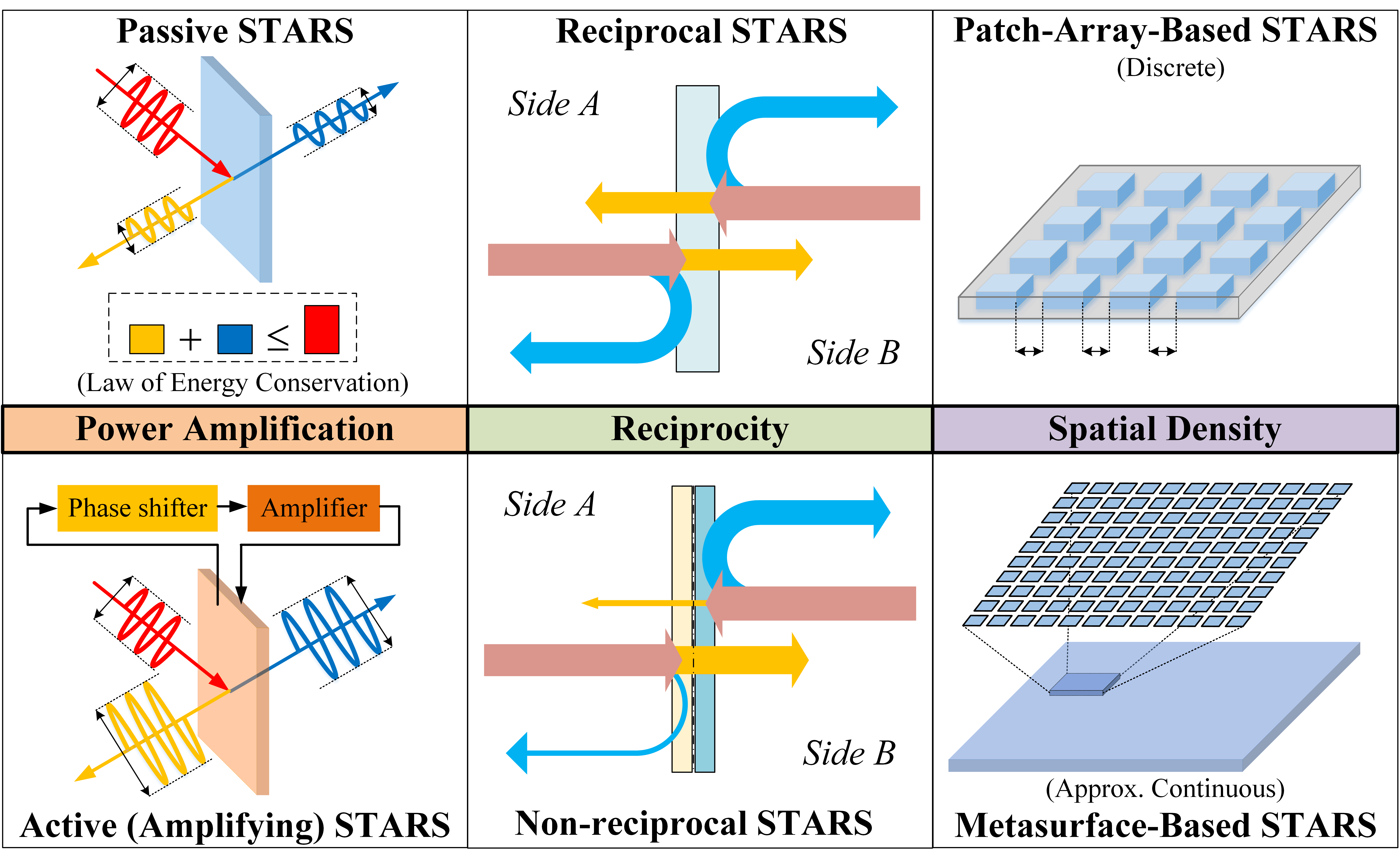}
    \caption{Different types of STARS employed in wireless communications.}
    \label{STAR_2}
\end{center}
\end{figure*}

\section{STARS for Communications}
In this section, we first introduce the different categories of STARS employed in wireless communication in terms of three characteristics, namely power amplification ability, reciprocity, and element spatial density, as illustrated in Fig. \ref{STAR_2}.

\subsection{Passive and Active STARS}
To begin with, we introduce passive STARS and active STARS, which are distinguished based on whether the incident wireless signals can be amplified or not.
\subsubsection{Passive STARS} Passive STARS mainly relies on low-cost EM elements, such as positive-intrinsic-negative (PIN) diodes and varactors, with negligible power consumption. By changing the states of these elements, passive STARS is only able to manipulate the incident wireless signal subject to the given signal's energy, i.e., following the law of energy conservation. In the top left of Fig. \ref{STAR_2}, it can be observed that, for passive STARS, the sum of transmitted and reflected signals' energy cannot exceed the energy of the incident signal. Therefore, passive STARS has to strike a trade-off between the signal quality for both sides.
\begin{itemize}	
	\item \emph{Pros}: The most attractive part of passive STARS is the cost- and energy-efficient feature, which is of vital importance for network operators who suffer from the high investment cost and long-term power cost for building and maintaining 6G. The economically viable passive STARS can incorporate thousands of elements, where the far-field signal-to-noise ratio (SNR) scaling law increases quadratically with the number of passive STARS elements~\cite{10163896}. Moreover, given the enlarged aperture size, passive STARS will result in significantly large near-field regions, which provide new design opportunities for communications and sensing~\cite{10558818}.
    \item \emph{Cons}: Due to the lack of amplifiers and radio-frequency (RF) chains, the coverage of passive STARS is generally limited due to the long-distance path loss. Thus, passive STARS is only beneficial for wireless nodes located in its vicinity. Moreover, the acquisition of channel state information (CSI) for both transmitting and reflecting links is a challenging task since passive STARS does not have the signal processing capability.
\end{itemize}
\begin{figure}[t!]
\begin{center}
    \includegraphics[width=3in]{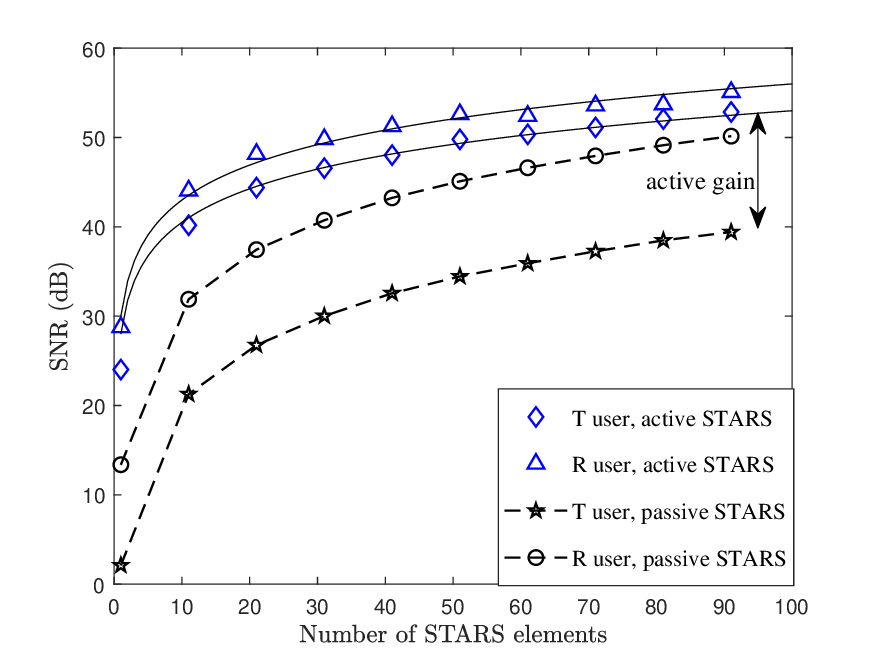}
    \caption{SNR scaling laws achieved by active and passive STARS for transmitted/reflected (T/R) users. The simulation parameters can be found in~\cite{10163896}.}
    \label{STAR_3}
\end{center}
\end{figure}
\subsubsection{Active STARS} Active STARS further incorporates amplifiers with dedicated power suppliers, as shown at the bottom left of Fig. \ref{STAR_2}. As a result, the amplitudes of both transmitted and reflected signals can be further enlarged and even larger than that of the incident signal. According to the hardware models proposed in~\cite{10163896}, if only one reflection-type amplifier is installed in the active STARS element, the amplitude controls for both sides are identical. However, independent amplitude control can be achieved with the use of two reflection-type amplifiers.
\begin{itemize}	
	\item \emph{Pros}: By leveraging its amplification capabilities, active STARS can mitigate the ``double-fading effect'' encountered by passive STARS, thus enhancing wireless coverage. Furthermore, when equipped with RF chains, active STARS can effectively address CSI estimation challenges and unlock a range of advanced functionalities, such as sensing and computing.
    \item \emph{Cons}: Besides the high energy consumption issue, active STARS introduces additional noise during the amplification. As a result, the corresponding far-field SNR scaling law achieved by active STARS only increases linearly with the number of elements~\cite{10163896}.
\end{itemize}

\subsubsection{Numerical Examples} Fig. \ref{STAR_3} depicts the scaling law of the SNR achieved by passive and active STARS with independent transmission and reflection control. As expected, the quadratic scaling law is achieved by the passive STARS, while the linear scaling law is observed for the active STARS. Moreover, it can be observed that the active gain is more pronounced when there is a limited STARS element. However, with the increasing of the STARS size, passive STARS has the potential to outperform active STARS since there is no additional noise introduced by amplifiers.

\subsection{Reciprocal and Non-reciprocal STARS}
In terms of reciprocity, we can also classify STARS into reciprocal STARS and non-reciprocal STARS.

\subsubsection{Reciprocal STARS} For reciprocal STARS, the pair of transmission and reflection coefficients is isotropic. As illustrated in the top middle of Fig. \ref{STAR_2}, the wireless signals incident upon both sides of reciprocal STARS have to follow identical signal manipulation for transmission and reflection. In other words, there is only a single pair of transmission and reflection coefficients that can be adjusted in reciprocal STARS, regardless of which sides the wireless signals come from.  
\begin{itemize}	
	\item \emph{Pros}: Reciprocal STARS is generally easy to implement with fewer constraints on the employed metamaterials. Moreover, reciprocal STARS retains the channel reciprocity for carrying out the CSI estimation in time division duplex systems. 
    \item \emph{Cons}: An inherent challenge with reciprocal STARS is the leakage of signal energy in uplink communication scenarios. For instance, when users surrounding the reciprocal STARS send data to the base station (BS) located on one side, a portion of the signal inevitably will transmit or reflect into the other side without the BS, thus constraining energy efficiency. To mitigate this issue, reciprocal STARS can periodically work in the pure transmission and pure reflection modes to fully forward the users' signals from each side in a successive manner. This, however, increases the control overheads.
\end{itemize}

\subsubsection{Non-reciprocal STARS} By contrast, non-reciprocal STARS provides more DoFs, where the transmission and reflection coefficients on both sides can be different, as depicted in the bottom middle of Fig. \ref{STAR_2}. As a result, independent transmission and reflection signal manipulations can be achieved for incident wireless signals on each side.
\begin{itemize}	
	\item \emph{Pros}: Non-reciprocal STARS incorporates reciprocal STARS as a special case and therefore provides more DoFs for communication designs. The uplink energy loss issue can be mitigated with properly designed transmission and reflection coefficients on both sides, e.g., a pure transmission mode on one side while a pure reflection mode on the other side.
    \item \emph{Cons}: Nevertheless, non-reciprocal STARS has complex hardware designs (e.g., multi-layer structures) and critical requirements on metamaterials. 
\end{itemize}

\subsection{Patch-array-based and Metasurface-based STARS}
The physical size of STARS elements is another important characteristic, which determines the maximum spatial density of STARS elements. Accordingly, we have patch-array-based STARS and metasurface-based STARS.
\subsubsection{Patch-array-based STARS} For patch-array-based STARS, the EM elements usually employ PIN diodes and varactors, with both the physical size and element spacing typically measured in the order of a few centimeters. Consequently, for a given aperture size, patch-array-based STARS exhibits a low element spatial density, as shown in the top right of Fig. \ref{STAR_2}. Additionally, considering that STARS elements are typically spaced at half-wavelength intervals to generate the desired transmitting and reflecting beams, patch-array-based STARS is mainly employed at low operating frequencies.
\begin{itemize}	
	\item \emph{Pros}: The low element spatial density makes patch-array-based STARS easy to be fabricated and configured. Another advantage of patch-array-based STARS is the low-cost feature, thus being widely investigated for wireless communications.
    \item \emph{Cons}: Patch-array-based STARS typically cannot be utilized for mm-wave and THz communications due to the small half-wavelength intervals, which cannot accommodate these EM elements effectively.
\end{itemize}

\subsubsection{Metasurface-based STARS} As depicted in the bottom right of Fig. \ref{STAR_2}, the dimension of elements in metasurface-based STARS is significantly reduced to the order of a few millimeters or even to the molecular scale. This characteristic endows metasurface-based STARS with an approximated continuous structure.
\begin{itemize}	
	\item \emph{Pros}: Compared to patch-array-based STARS, the massive number of elements and approximated continuous structure enable metasurface-based STARS to have much higher spatial DoFs for capacity enhancement~\cite{10232975}.
    \item \emph{Cons}: The hardware design and configuration of metasurface-based STARS become quite complicated. Additionally, the approximated continuous structure requires new channel modelling and beamforming design for operating metasurface-based STARS in wireless communications.
\end{itemize}

\section{STARS for Sensing}
Sensing is a crucial feature of 6G wireless networks, designed to deliver environment-aware services for emerging applications like vehicle-to-everything, smart homes, and smart cities, as well as enhancing basic wireless communication performance. Specifically, sensing is to acquire data about passive targets, such as their positions, velocities, and orientations, by transmitting probing signals and analyzing the echoes reflected by these targets. Unlike wireless communication, which benefits from non-line-of-sight (NLoS) propagation for signal multiplexing, wireless sensing primarily depends on the line-of-sight (LoS) path between the sensing node and targets due to the lack of reference points in NLoS paths. RISs are seen as a promising solution to overcome LoS blockage issues in sensing systems by creating a virtual LoS path and serving as an additional reference point \cite{wymeersch2020radio}. STARS inherits all the advantages of reflective/transmissive RISs for sensing and offers extra benefits due to its full-space properties. Next, we will introduce basic sensing architectures utilizing STARS and discuss the corresponding key benefits and challenges.

\begin{figure*}[t!]
\begin{center}
    \includegraphics[width=0.9\textwidth]{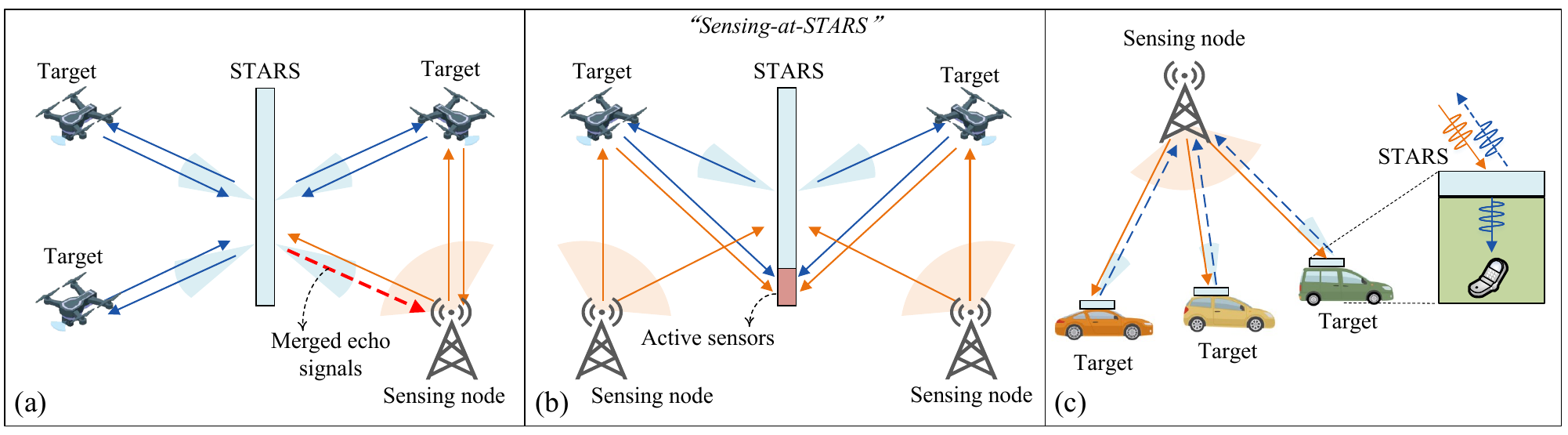}
    \caption{Conceptual illustration of three proposed sensing architecture utilizing STARS: (a) (a) STARS-aided monostatic sensing, where the transmission sensing node also serves as the reception node; (b) STARS-enabled bistatic sensing, where STARS functions as the reception node with the aid of active sensors; and (c) Sensing with target-mounted STARS, where STARS controls signal reflection at the target and enhances the communication service inside the target.}
    \label{fig:STARS_sensing}
\end{center}
\end{figure*}

\subsection{STARS-aided Monostatic Sensing}

Monostatic sensing refers to a sensing system with collocated transmitter and receiver. 
Fig. \ref{fig:STARS_sensing}(a) depicts the architecture of a STARS-aided monostatic sensing system, with a single sensing node and multiple passive targets distributed across two separate spaces. As demonstrated, the STARS not only establishes a virtual LoS path and serves as an additional reference point for sensing targets on the same side as the sensing node, but it also relays the probing signal to the opposite side, thereby expanding the coverage area. 

\begin{itemize}
    \item \emph{Benefits}: In monostatic sensing, STARS can be seamlessly integrated as an add-on solution, enabling the sensing node to monitor targets across two separate spaces. This direct integration offers the advantage of minimizing additional hardware costs for existing systems.

    \item \emph{Challenges}: STARS-aided monostatic sensing presents two primary challenges. Firstly, the sensing node receives merged echo signals from targets on different sides, complicating the task of distinguishing between the targets and determining their specific locations using conventional low-complexity sensing algorithms such as the sparsity-based compressed sensing method. Secondly, the echo signal strength at the sensing node is often weak due to significant multiplicative path loss from the multiple hops between the sensing node, STARS, and targets, as depicted in Fig. \ref{fig:STARS_sensing}(a). Additionally, the signal undergoes double-time energy splitting at the STARS, further diminishing its strength. Active STARS and/or non-reciprocal STARS can be employed to address the issue of low signal strength and signal leakage, which, however, introduces additional hardware costs and poses challenges in beamforming design.
\end{itemize}

\subsection{STARS-enabled Bistatic Sensing: Sensing-at-STARS}

Unlike monostatic sensing, bistatic sensing utilizes sensing nodes located at different positions for transmission and reception. In this arrangement, the placement of the reception node is crucial for effective sensing performance, particularly when employing the STARS and considering that the target may be positioned on different sides relative to the STARS. If the reception node is situated on one specific side of the STARS, bistatic sensing faces similar challenges to monostatic sensing in terms of low signal strength caused by multiple hops of signals and the high-complexity sensing algorithms caused by merged echo signals. To address these issues and fully reap the benefits of bistatic sensing in STARS systems, a \emph{sensing-at-STARS} architecture has been recently proposed \cite{wang2023stars}, as shown in Fig. \ref{fig:STARS_sensing}(b). In this design, active sensors are mounted on the STARS, allowing STARS to function as the reception node in bistatic sensing.

\begin{itemize}
    \item \emph{Benefits}: Sensing-at-STARS effectively addresses the key challenges encountered in STARS-aided monostatic sensing. In this architecture, active sensors mounted on the STARS receive echo signals directly from the targets, ensuring that the signals are not merged. Under this circumstance, it is easy to identify which side the targets are located on and the design of the sensing algorithm can be significantly simplified. By reducing the number of signal hops from the transmitter to the receiver, the strength of the echo signals is considerably enhanced, enhancing the sensing accuracy. Additionally, compared to reflective/transmissive RISs, the sensing-at-STARS architecture also enables efficient cooperative sensing between the sensing nodes located on different sides through the transmission and reflection beamforming as well as the joint reception and processing of all echo signals from different nodes at the active sensors, reducing the complexity of interference management and data fusion.

    \item \emph{Challenges}: However, sensing-at-STARS presents two primary challenges. First, the integration of RF chains and signal-processing modules at the STARS is required to receive and analyze echo signals, which inevitably raises hardware costs. Second, discrepancies in the clock timing between the active sensors and the sensing node can cause offsets in the time, frequency, and phase of the echo signal, thereby diminishing the accuracy of the sensing. Consequently, it is crucial to ensure rigorous clock synchronization between the active sensors and the sensing node or, alternatively, to develop sensing algorithms that are aware of and can compensate for synchronization errors.
\end{itemize}

\begin{figure*}[t!]
\begin{center}
    \includegraphics[width=5in]{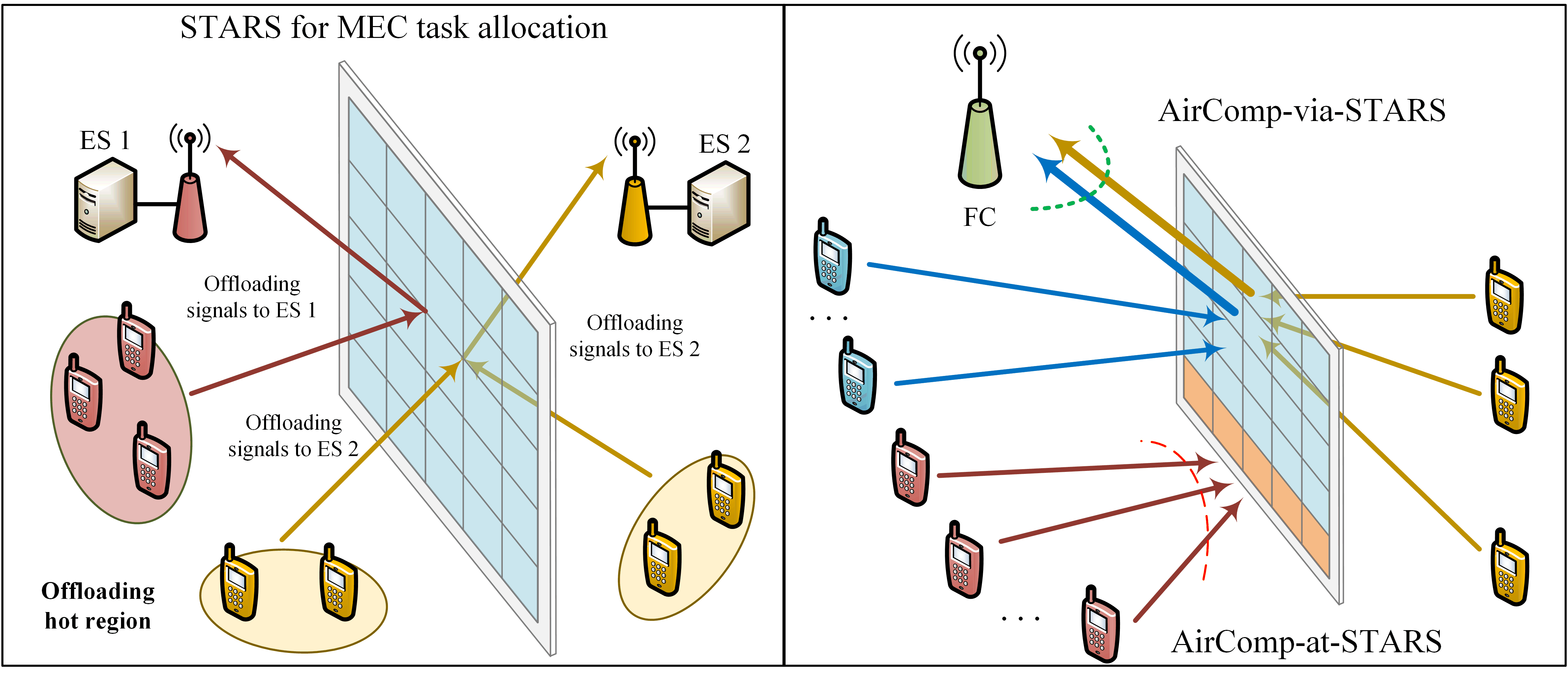}
    \caption{STARS for Computing in 6G.}
    \label{STAR_4}
\end{center}
\end{figure*}

\subsection{Sensing with Target-mounted STARS }
For either monostatic or bistatic sensing, the signal reflection at the target is typically random and can change rapidly in high-mobility scenarios. Apart from exploiting STARS to establish virtual LoS paths and enlarge the coverage area for sensing, we can also realize controllable reflection of signals by mounting STARS on the target \cite{meng2023sensing}, as illustrated in Fig. \ref{fig:STARS_sensing}(c). In particular, the echo signal from the target can be crafted through the reflection beamforming of STARS. Moreover, the signal impinged on the target can also go through the STARS and provide additional communication service to the devices inside the target (e.g., mobile users in a vehicle), which is challenging for conventional reflective/transmissive RISs.

\begin{itemize}
    \item \emph{Benefits}: Through reflection beamforming, the echo signal can be dynamically adjusted to achieve maximum power directed toward the sensing node. This ensures that a stable and strong echo signal is consistently received at the sensing node, while also reducing the likelihood of detrimental multipath effects. Moreover, the incorporation of target-mounted STARS also facilitates the integration of sensing and communication. Specifically, by boosting the echo signal's strength, the sensing node can effectively track the target and acquire CSI in a pilot-free manner. This enables efficient communication to reliably convey information to the mobile user located within the target using STARS transmission beamforming.

    \item \emph{Challenges}: Establishing a control link between the sensing node and the target-mounted STARS is challenging due to the non-fixed position of the target. Therefore, a new control protocol for the STARS may be needed to facilitate the beamforming design.
\end{itemize}

\section{STARS for Computing and Caching}
Having discussed the employment of STARS for wireless sensing, in this section, we focus our attention on STARS for supporting computing and caching functionalities with wireless networks.

\subsection{STARS for Computing}
High-performance computing plays an important role in the success of artificial intelligence (AI) and its wide applications. Ubiquitous wireless networks allow connected devices to access expensive computing resources anywhere and anytime for facilitating mobile AI applications and services. Existing work has investigated the employment of multiple transmissive metasurfaces, namely stacked intelligent metasurfaces (SIMs), to act as a diffractive neural network to carry out computation tasks during wave propagation \cite{10515204}. In the following, we will discuss the promising applications of STARS for further boosting the computing functionality in 6G wireless networks.
\subsubsection{STARS-aided MEC}
MEC stands as a pivotal enabling technology for mobile AI, where mobile devices with limited onboard computing resources offload computational tasks (e.g., machine learning training tasks) to edge servers (ESs) via wireless connections~\cite{8016573}. As shown in the left side of Fig. \ref{STAR_4}, compared to reflective/transmissive RISs, STARS not only grants access to computing services for all mobile devices on both sides but also can achieve dynamic task offloading to any ESs located on either side. As depicted, computational tasks from numerous mobile devices in the offloading hot region can be efficiently distributed, with a portion allocated to ESs that are either idle or less occupied on the opposite side. This unique flexibility provided by STARS-aided MEC can further improve the computation efficiency and thus reduce the computation delays.

\subsubsection{STARS-aided AirComp} AirComp is another wireless computing technology, which is distinct from MEC's `computing-after-communication' protocol, mobile devices in AirComp send pre-processed signals to the destination (namely fusion center (FC)) and FC directly processes the superimposed signals for the computational result, thus providing the concept of `computing-while-communication'~\cite{9095231}. STARS provides the following two operations for AirComp, as shown in the right side of Fig. \ref{STAR_4}.

\begin{itemize}	
	\item \emph{AirComp-via-STARS}: For AirComp, the pre-processed signal of each mobile user is essential for final computational results. The `AirComp-via-STARS' operation exploits the full-space coverage of STARS and ensures that each mobile user's signal can be redirected to the FC. This is generally impossible for reflective/transmissive RISs.
    \item  \emph{AirComp-at-STARS}: In the `AirComp-at-STARS' operation, where RF chains are linked to certain elements of STARS, pre-processed signals from mobile users on both sides are initially aggregated at the STARS to mitigate long-distance path loss. Subsequently, the AirComp result can be transmitted to the FC via the wired link. However, to efficiently operate the `AirComp-at-STARS', appropriate protocols have to be developed to coordinate the fusion of specific signals at the FC or STARS and mitigate the resulting interference caused by undesired signals.
\end{itemize}

\subsection{Caching-at-STARS}
With the advancement of the Metaverse and digital twin technologies, there is a growing demand for delivering vast amounts of Internet-based content, such as videos and images, to end-users over wireless networks while maintaining minimal latency. To this end, edge caching has emerged as a solution to alleviate network throughput demands. In particular, edge caching stores popular content at network ESs situated near end-users, thereby enhancing accessibility and reducing latency~\cite{8676308}. To unlock the potential of STARS for edge caching, a novel `caching-at-STARS' architecture was proposed and examined in \cite{10388256}. As depicted in the middle top of Fig. \ref{STAR_1}, caching-at-STARS equips with cache units at the STARS. On the one hand, the traditional edge caching links between ESs and end-users on both sides can be enhanced via the STARS. On the other hand, caching units deployed at the STARS offer a closer caching service to end-users compared to ESs, further improving content delivery efficiency and reducing latency. Moreover, widely deployed STARS with full-space coverage provide a significant distributed caching space for wireless networks compared to reflective/transmissive RISs with limited coverage. This extensive deployment enhances caching capabilities, enabling efficient storage and retrieval of content across networks, ultimately improving the overall user experience.

\begin{figure}[t!]
\begin{center}
    \includegraphics[width=3in]{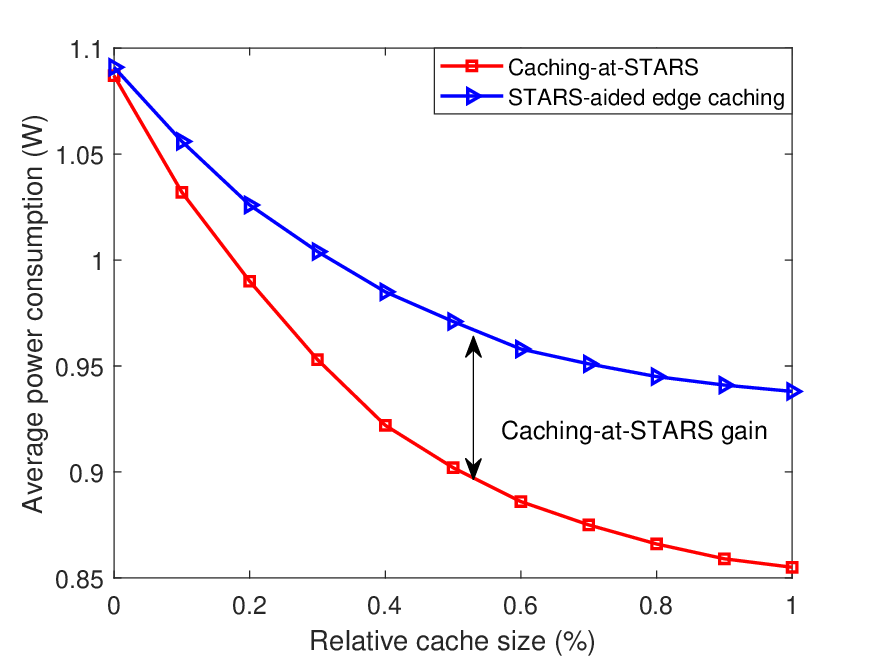}
    \caption{Average network power consumption versus the relative cache size achieved by `caching-at-STARS' and `STARS-aided edge caching' schemes. The simulation parameters can be found in \cite{10388256}.}
    \label{STAR_5}
\end{center}
\end{figure}

To further demonstrate the advantages of the caching-at-STARS architecture, Fig. \ref{STAR_5} depicts the average network power consumption versus the relative cache size when employing one STARS in edge caching networks. Here, the relative cache size is defined as the proportion of the cache size allocated at the STARS relative to the total size of content across the entire network. The cache size allocated at the ES is presumed to be twice that of the STARS. It can be observed that caching-at-STARS can significantly reduce power consumption compared to the conventional STARS-aided edge caching scheme. The performance gain becomes more significant when the STARS's cache size increases since caching-at-STARS has a shorter distance to end-users. The above results confirmed the effectiveness of caching-at-STARS. 

\section{State of the Art for Standardization and Outlook}
There has been a noticeable increase in standardization activities concerning the integration of RIS in next-generation wireless networks, such as the International Telecommunication Union (ITU) and European Telecommunications Standards Institute (ETSI). In particular, in the report titled ``Framework and Overall Objectives of the Future Development of IMT for 2030 and Beyond'' published by ITU in November 2023, RISs are identified as one of the potential technologies to improve communication performance and overcome challenges in traditional beam-space antenna array beamforming. In the context of ETSI, it launched a dedicated industry specification group (ISG) in September 2021 to study and discuss the standardization of RIS for 6G. Note that although RIS has not been formally studied in the 3rd Generation Partnership Project (3GPP), a study item on the network-controlled repeater (NCR) was initiated in Release 18. Here, NCR is an enhancement over conventional RF repeaters with the capability to receive and process side control information from the network, which shares some common characteristics with RISs. In the following, we focus our attention on the completed and ongoing activities of ETSI-ISG-RIS~\cite{ETSI-ISG-RIS}, while discussing the correlations and implications for studying STARS in multi-functional 6G wireless networks.

There were 3 work items (WIs) from 2021 to 2023, which are outlined as follows.
\begin{itemize}	
	\item \emph{WI-1: Use Cases, Deployment Scenarios and Requirements}: The concept of STARS was initially raised in this WI to combine the reflection and refraction into one RIS. The use cases, deployment scenarios, and operational requirements identified here provide important references for the application of STARS in multi-functional 6G.
    \item \emph{WI-2: Technological Challenges, Architecture, and Impact on Standardization}: The diverse controlling mechanism discussed here can be used to guide how to control STARS to support multiple functionalities. Moreover, this WI also suggests different topologies for facilitating integrated sensing and communications with RIS, which can be further considered with STARS.  
    \item \emph{WI-3: Communication Models, Channel Models, Channel Estimation, and Evaluation Methodology}: This WI provides a comprehensive discussion on channel modelling associated with RIS for both communications and sensing functionalities, which can be reused for patch-array-based STARS. 
\end{itemize}

Currently, there are 4 ongoing WIs since September 2023, which are discussed as follows.

\begin{itemize}	
	\item \emph{WI-4: Implementation and Practical Considerations}: At the time of writing, this WI is close to a stable version. In this WI, hardware design and practical implementation for reflective, refractive, and absorptive RISs were presented. This guides the development of STARS prototypes and their practical implementation. 
    \item \emph{WI-5: Diversity and Multiplexing of RIS-aided Communications}: This WI aims to reveal the diversity and multiplexing performance of RIS across various domains. The fundamental performance limits and DoFs identified in this WI can serve as valuable references for designing interference management strategies in STARS-aided multi-functional 6G.
    \item \emph{WI-6: Multi-functional RIS: Modelling, Optimisation, and Operation}: This WI incorporates STARS as one type of multi-functional RISs and will also explore the potential utilization of RIS with sensing and computing functionalities. These ongoing studies will offer valuable insights for STARS-aided multi-functional 6G.
    \item \emph{WI-7: Near-field Channel Modelling and Mechanics}: This WI aims to concentrate specifically on near-field channel modelling for RISs. The enhanced DoFs and novel designs enabled by near-field channels can further benefit STARS-aided multi-functional 6G.
\end{itemize}

\section{Conclusions}
In this article, the employment of STARS for multi-functional 6G wireless networks has been discussed. In particular, different types of STARS used in communications were introduced from the perspectives of power amplification capability, reciprocity, and elements' spatial density, where the corresponding advantages and disadvantages were discussed. Furthermore, three wireless sensing architectures based on STARS, namely STARS-aided monostatic, STARS-enabled bistatic sensing, and sensing with target-mounted STARS, were proposed. For each sensing architecture, promising benefits and application challenges were discussed. Moreover, several applications of STARS for MEC, Aircomp, and edge caching were put forward to support the computing and caching functionalities in wireless networks. Finally, recent standardization activities of RISs have been outlined and their possible contributions to facilitating STARS-aided multi-functional 6G were highlighted.

\bibliographystyle{IEEEtran}
\bibliography{mybib}

\begin{thebibliography}{10}
\providecommand{\url}[1]{#1}
\csname url@samestyle\endcsname
\providecommand{\newblock}{\relax}
\providecommand{\bibinfo}[2]{#2}
\providecommand{\BIBentrySTDinterwordspacing}{\spaceskip=0pt\relax}
\providecommand{\BIBentryALTinterwordstretchfactor}{4}
\providecommand{\BIBentryALTinterwordspacing}{\spaceskip=\fontdimen2\font plus
\BIBentryALTinterwordstretchfactor\fontdimen3\font minus \fontdimen4\font\relax}
\providecommand{\BIBforeignlanguage}[2]{{%
\expandafter\ifx\csname l@#1\endcsname\relax
\typeout{** WARNING: IEEEtran.bst: No hyphenation pattern has been}%
\typeout{** loaded for the language `#1'. Using the pattern for}%
\typeout{** the default language instead.}%
\else
\language=\csname l@#1\endcsname
\fi
#2}}
\providecommand{\BIBdecl}{\relax}
\BIBdecl

\bibitem{10529727}
A.~Kaushik, R.~Singh, S.~Dayarathna, R.~Senanayake, M.~Di~Renzo, M.~Dajer, H.~Ji, Y.~Kim, V.~Sciancalepore, A.~Zappone, and W.~Shin, ``Toward integrated sensing and communications for 6{G}: Key enabling technologies, standardization, and challenges,'' \emph{IEEE Commun. Stand. Mag.}, vol.~8, no.~2, pp. 52--59, June 2024.

\bibitem{9140329}
M.~Di~Renzo, A.~Zappone, M.~Debbah, M.-S. Alouini, C.~Yuen, J.~de~Rosny, and S.~Tretyakov, ``Smart radio environments empowered by reconfigurable intelligent surfaces: How it works, state of research, and the road ahead,'' \emph{{IEEE} J. Sel. Areas Commun.}, vol.~38, no.~11, pp. 2450--2525, Nov. 2020.

\bibitem{9570143}
X.~Mu, Y.~Liu, L.~Guo, J.~Lin, and R.~Schober, ``Simultaneously transmitting and reflecting ({STAR}) {RIS} aided wireless communications,'' \emph{{IEEE} Trans. Wireless Commun.}, vol.~21, no.~5, pp. 3083--3098, May 2022.

\bibitem{10163896}
J.~Xu, J.~Zuo, J.~T. Zhou, and Y.~Liu, ``Active simultaneously transmitting and reflecting ({STAR})-{RIS}s: Modeling and analysis,'' \emph{{IEEE} Commun. Lett.}, vol.~27, no.~9, pp. 2466--2470, Sept. 2023.

\bibitem{10558818}
J.~An, C.~Yuen, L.~Dai, M.~Di~Renzo, M.~Debbah, and L.~Hanzo, ``Near-field communications: Research advances, potential, and challenges,'' \emph{{IEEE} Wireless Commun.}, vol.~31, no.~3, pp. 100--107, 2024.

\bibitem{10232975}
T.~Gong, P.~Gavriilidis, R.~Ji, C.~Huang, G.~C. Alexandropoulos, L.~Wei, Z.~Zhang, M.~Debbah, H.~V. Poor, and C.~Yuen, ``Holographic {MIMO} communications: Theoretical foundations, enabling technologies, and future directions,'' \emph{{IEEE} Commun. Surv. Tut.}, vol.~26, no.~1, pp. 196--257, 2024.

\bibitem{wymeersch2020radio}
H.~Wymeersch, J.~He, B.~Denis, A.~Clemente, and M.~Juntti, ``Radio localization and mapping with reconfigurable intelligent surfaces: Challenges, opportunities, and research directions,'' \emph{{IEEE} Veh. Technol. Mag.}, vol.~15, no.~4, pp. 52--61, Dec. 2020.

\bibitem{wang2023stars}
Z.~Wang, X.~Mu, and Y.~Liu, ``{STARS} enabled integrated sensing and communications,'' \emph{{IEEE} Trans. Wireless Commun.}, vol.~22, no.~10, pp. 6750--6765, Oct. 2023.

\bibitem{meng2023sensing}
K.~Meng, Q.~Wu, W.~Chen, and D.~Li, ``Sensing-assisted communication in vehicular networks with intelligent surface,'' \emph{{IEEE} Trans. Veh. Technol.}, vol.~73, no.~1, pp. 876--893, Jan. 2024.

\bibitem{10515204}
J.~An, C.~Yuen, C.~Xu, H.~Li, D.~W.~K. Ng, M.~D. Renzo, M.~Debbah, and L.~Hanzo, ``Stacked intelligent metasurface-aided {MIMO} transceiver design,'' \emph{{IEEE} Wireless Commun.}, Early Access, doi: 10.1109/MWC.013.2300259.

\bibitem{8016573}
Y.~Mao, C.~You, J.~Zhang, K.~Huang, and K.~B. Letaief, ``A survey on mobile edge computing: The communication perspective,'' \emph{IEEE Commun. Surveys Tuts.}, vol.~19, no.~4, pp. 2322--2358, Fourthquarter 2017.

\bibitem{9095231}
W.~Liu, X.~Zang, Y.~Li, and B.~Vucetic, ``Over-the-air computation systems: Optimization, analysis and scaling laws,'' \emph{{IEEE} Trans. Wireless Commun.}, vol.~19, no.~8, pp. 5488--5502, Aug. 2020.

\bibitem{8676308}
J.~Yao, T.~Han, and N.~Ansari, ``On mobile edge caching,'' \emph{IEEE Commun. Surveys Tuts.}, vol.~21, no.~3, pp. 2525--2553, Thirdquarter 2019.

\bibitem{10388256}
Z.~Hu, R.~Zhong, C.~Fang, and Y.~Liu, ``Caching-at-{STARS}: the next generation edge caching,'' \emph{{IEEE} Trans. Wireless Commun.}, vol.~23, no.~8, pp. 8372--8387, 2024.

\bibitem{ETSI-ISG-RIS}
\BIBentryALTinterwordspacing
``{ETSI-ISG-RIS} work programme''. [Online]. Available: \url{https://portal.etsi.org/tb.aspx?tbid=900#/lt-50611-work-programme}
\BIBentrySTDinterwordspacing

\end{thebibliography}






\end{document}